\def\pmb#1{\setbox0=\hbox{#1}%
  \kern-.025em\copy0\kern-\wd0
  \kern.05em\copy0\kern-\wd0
  \kern-.025em\raise.0433em\box0}

\def\bftheta{\pmb{$\theta$}}

\documentstyle[aps,preprint]{revtex}
\topmargin = - 0.9 cm
\begin{document}
\begin{titlepage}
\begin{flushright}
MIT-CTP\#2324\\
IFUP-TH-33/94
\end{flushright}
\vskip 1truecm
\begin{center}
\Large\bf
Energy Theorem for 2+1 dimensional gravity\footnote{This work is  supported in
part by D.O.E. cooperative agreement DE-FC02-94ER40818 and by M.U.R.S.T.}
\end{center}
\medskip
\begin{center}
Pietro Menotti\\
{\small\it
Dipartimento di Fisica dell' Universit\'a, Pisa 56100, Italy and\\[-5.5pt]
I.N.F.N., Sezione di Pisa.}\\[-4pt]
{\small \tt e-mail: menotti@ibmth.difi.unipi.it}\\
{\small\rm  and}\\
Domenico Seminara\\
{\small\it
Center~for~Theoretical~Physics,~{Laboratory~for~Nuclear~Science~and~Department~of~Physics}\footnote{
Present address}\\[-4pt]
Massachusetts Institute of Technology, Cambridge, MA 02139 U.S.A.,}\\[-4pt]
\small\it
Scuola Normale Superiore, Pisa 56100, Italy, and
I.N.F.N., Sezione di Pisa.\\ [-4pt]
{\small \tt e-mail: seminara@pierre.mit.edu}
\end{center}
\begin{center}
Submitted to: Annals of Physics
\end{center}
\date{June 1994}

\end{titlepage}

\begin{abstract}
We prove a positive energy theorem in 2+1 dimensional gravity for open
universes and any matter energy-momentum tensor satisfying the
dominant
energy condition. We consider on the  space-like initial value surface
a family of widening Wilson loops and show that the energy-momentum of
the enclosed subsystem is a future directed time-like vector whose
mass
is an increasing function of the loop, until it
reaches the value $1/4G$ corresponding to a deficit angle of
$2\pi$.
At this point the energy-momentum of the system evolves, depending on
the nature of a zero norm vector appearing in the evolution equations,
either into a time-like vector of a universe which closes
kinematically or into a Gott-like universe whose energy momentum
vector, as first recognized by Deser, Jackiw and 't Hooft is space-like.
 This treatment generalizes results obtained
by Carroll, Fahri, Guth and Olum  for a
system of point-like spinless particle, to the most general form of
matter whose energy-momentum tensor satisfies the dominant energy
condition.
The treatment is also given for the anti de Sitter 2+1 dimensional gravity.

\end{abstract}
\section{Introduction.}
The energy-momentum and angular momentum of a system in general
relativity in $n+1$
dimensions with $n\ge 3$ is defined by means of the flux of a
pseudotensor through a $n-1$ dimensional space-like surface located at
space infinity. It is well known that such energy-momentum can be
defined when the metric approaches at space infinity the minkowskian
metric \cite{ADMLW}.
Such a definition cannot be taken over directly to $2+1 $ dimensional
gravity; in fact it is true that the space outside the sources is
exactly flat, but the global structure of the space time at space
infinity is that of a cone with a non zero deficit angle. In \cite{DJH}
 it is proved that for a
stationary, static distribution of matter such a deficit angle
equals the space integral of ${\cal T}^{0}_0$  i.e. the total amount of mass of
the matter present in the system. Such result however is no longer
true in the general situation. With regard to the angular momentum the
situation in more critical; one can consider a rather artificial
situation in which the deficit angle is zero in which case one can
carry over the usual procedure of 3+1 dimension and thus one is able
\cite{DJH}
in the stationary case to identify the angular momentum with the time
jump which occurs in a syncronous reference system when one performs a
closed trip around the source.

A more general definition of energy-momentum and angular momentum in
2+1 dimension can be derived by computing the holonomies of the
$ISO(2,1)$ group which were introduced in connection to the
formulation of pure 2+1 dimensional gravity as a Chern-Simon theory
\cite{Witten}.
Such idea of defining mass and angular momentum  in 2+1 dimensional gravity
by means of holonomies is already contained in the seminal paper \cite{DJH}
for a system of point-like spinless particles.
{}From the Lorentz holonomy one can extract a three component vector
which transforms like a Lorentz vector under local Lorentz
transformations. Such a vector does not depend on deformations of
the loop keeping the origin fixed provided
such loop extends only outside the matter, and its square can be
naturally identified with the square of the mass of the system. Change
in the origin of the loop is equivalent to a Lorentz transformation
on such a vector. The above described definition of
the energy-momentum of a system in 2+1 dimensional gravity
has been applied in \cite{CFGO}
for analyzing general properties of a system of spinless point
particles which form an open universe. The important result of their
investigation is that if a subsystem of such a collection of particles
has space-like momentum the whole system must possess total space-like
momentum; in particular if a Gott pair exists in the system,
necessarily the universe is either closed, or if open it possesses a
space-like momentum.

Here adopting the same definition of the energy-momentum
we shall derive general properties for a system whose
matter energy-momentum tensor satisfies the dominant energy condition
(DEC). We recall that the dominant energy condition \cite{HawEll}  states
that for any time-like future directed vector $v^\mu$ its contraction with the
energy momentum tensor $T^\mu_\nu v^\nu$ is time-like future directed.
We shall prove a positive energy theorem
in 2+1 dimension which can be stated as follows : Given a space-like
initial data two dimensional surface consider on it a family of expanding
Wilson loops; if the matter energy-momentum tensor satisfies the
dominant energy condition (DEC), the energy-momentum of the subsystem
enclosed by the loop is a time-like future directed vector whose norm
increases as the loop embraces more and more matter. This monotonic
increase occurs until the mass of the system reaches the critical
value $1/4G$ corresponding to a deficit angle $2\pi$. After that point,
the evolution equations we derive in Sec.III show that two
alternatives can occur:
either the universe still increases its mass, going over to a closed
universe or it goes over to a Gott-like universe whose energy-momentum
is space-like. The choice depends on the nature of a vector
in the evolution  equation whose norm at the critical point is zero. If
such a vector is the zero vector then  the universe evolves into a
closed one; if it is light-like it goes over to the Gott-like
situation. Then we obtain again for a general matter  energy-momentum
tensor satisfying the DEC, the result of \cite{CFGO}
that once a subsystem of the universe has
space-like momentum by adding matter one can never recover an open
time-like universe.

{}From the $ISO(2,1)$ holonomy one can extract \cite{Witten} another scalar
invariant which is given by the scalar
product of the above described energy-momentum vector with another
three component object which obeys inhomogeneous
transformation properties. Such a second invariant
can be related to the total angular  momentum of the system which in 2+1
dimension has just one component.
Evolution  equation of similar type are also written for the angular
momentum.

As the situation in 2+1 dimensions is different from the one occurring
in higher dimensions, one needs to support such definitions of the
energy-momentum and angular momentum.

For the energy-momentum one can produce the following arguments 1) For
weak sources the given definition goes over the definition in
Minkowski space \cite{DJH}. 2) In the stationary static case the mass of
the system coincides with the sum of the masses of the sources (see
\cite{DJH} and Sect.III of the present paper). This is satisfactory because
in 2+1 dimensions there are no gravitational forces on thus no
gravitational potential. Thus one expects the above result in the
static case. 3) The theorem of Sect.III of the present paper
proves that if the sources satisfy the DEC and the total mass of
the system is less that the critical value $1/4G$ such vector in
addition to being conserved is time-like future directed.

With regard to the here adopted definition of angular momentum one
can produce the following arguments to support it 1) It is a conserved
quantity . 2) The vector whose scalar product with the energy-momentum
vector gives the angular momentum is subject to transformation
properties which are formally identical to those  that occur in
Minkowski space (see Appendix B of the present paper). 3) The solution
of the evolution equations for weak
sources gives for the angular momentum the same result as in Minkowski
space.

In Sect.V the described treatment is extended to 2+1
dimensional gravity in presence of a (negative) cosmological constant
i.e. the anti de Sitter case. Here due to the fact that the anti de
Sitter group is the direct product of two $SO(2,1)$ groups, one obtains
two independent equations similar to the one which intervene in the
Poincar\'e case and whose discussion to a large extent can be carried
through similarly to the previous case.

 \section{Energy in 2+1 dimensions}
The usual definition of energy which works in dimension equal or
higher than 3+1, does not apply to 2+1 dimensional gravity because
except for the empty space, the space is never asymptotically
minkowskian.  A definition of the energy in 2+1 dimensions similar to
the one given in 3+1 is obtained in \cite{BCJ} replacing the
asymptotic flat minkowski space with the conical space and computing
in carthesian coordinates
\begin{equation}
\label{E1}
\Theta_c=\frac{1}{2}\varepsilon_{abc}\oint \Gamma^{ab}_i dx^i;
\end{equation}
where $\Gamma^{ab}_\mu(x)$is the spin-connection and
the integral is evaluated on a loop at space infinity.  An alternative
expression equivalent to (\ref{E1}) which is fully covariant in the
sense that it can be computed in any coordinate system is provided by the
Lorentz holonomy
\begin{equation}
\label{E2}
W=P{\rm exp}\left [ -i \oint J_a \Gamma^a_\mu dx^\mu\right]= e^{-i J_a
\Theta^a }
\end{equation}
where $\Gamma^a_\mu=\frac{1}{2} \varepsilon^{a~~c}_{~~b}
\Gamma^b_{c\mu}$.  The Lorentz holonomy
 in (\ref{E2}) is along a closed contour but the trace is not
taken. The energy-momentum vector is defined by the $\Theta^a$ appearing on the
right hand side of (\ref{E2}) divided by $8\pi G $.
The reason why in conical coordinates (\ref{E2}) reduce to (\ref{E1})
is that in such coordinate system $\Gamma^a_\mu
dx^\mu=\displaystyle{8 \pi G M\delta^a_0  \frac{{\bf r}\wedge d{\bf r}}{{\bf
      r}^2}}$.
In the most general case the Lorentz holonomy cannot be written as a simple
exponential, but for example in the fundamental representation  on must allow
for a $\pm$ sign in front of it \cite{CFGO}.
The direction of $\Theta^a$ is identified by
the eigenvector (with eigenvalue 1) belonging to the adjoint
representation. Once the direction of $\Theta_a$ is established we can
distinguished three cases according to the norm of $\Theta^a$. In the
following we always use the fundamental representation of $SU(1,1)$ with
commutation relations given by
\begin{equation}
[ J_a, J_b]=i \epsilon_{abc}J^c
\end{equation}
and the traces given by
\begin{equation}
{\rm Tr}( J_a J_b)= -\frac{1}{2}\eta_{ab}
\end{equation}
where $\eta^{ab}\equiv {\rm diag}(-1,1,1)$.

{\noindent\it Case 1.}  $\Theta^a$ is a null vector. In this case the
expansion of (\ref{E2}) in the fundamental representation is given by
\begin{equation}
W=I- i \Theta^a J_a
\end{equation}
all the remaining terms being equal to zero  and taking the trace
with  $J^b$ one obtains
\begin{equation}
\Theta^b= - 2 i Tr( W J^b).
\end{equation}

{\noindent \it Case 2}. $\Theta^a$ is space-like.
\begin{equation}
W=\cosh\frac{\sqrt{\Theta^a \Theta_a}}{2} I-
2 i  \displaystyle{\frac{\sinh\sqrt{\Theta^a
      \Theta_a}/2}{\sqrt{\Theta^a \Theta_a}}}
(J_a \Theta^a).
\end{equation}
Taking the trace
\begin{equation}
 Tr(J^b W)= i \displaystyle{\frac{\sinh{\sqrt{\Theta^a
        \Theta_a}/2}}{\sqrt{\Theta^a \Theta_a}}} \Theta^b
\end{equation}
This determines completely $\Theta^b$  as $\sinh$ is an odd monotonic
function.

{\noindent \it Case 3}. $\Theta^a$ is time-like. The expansion in the
fundamental representation is
\begin{equation}
W=\cos \frac{\sqrt{-\Theta^a \Theta_a}}{2} I-
 2 i  \displaystyle{\frac{\sin\sqrt{-\Theta^a
       \Theta_a}/2}{\sqrt{-\Theta^a \Theta_a}}}
(J_a \Theta^a).
\end{equation}
Taking the trace
\begin{equation}
\label{AAA}
 Tr(J^b W)= i \displaystyle{\frac{\sin{\sqrt{-\Theta^a
        \Theta_a}/2}}{\sqrt{-\Theta^a \Theta_a}}} \Theta^b.
\end{equation}
Eq. (\ref{AAA}) is not sufficient to determine $\Theta^a$, but combining it
with the trace of $W$
\begin{equation}
Tr(W)=2 \cos\frac{\sqrt{-\Theta^a \Theta_a}}{2}
\end{equation}
$\Theta^a$ is fixed if one  imposes that $\delta\equiv 8 \pi G
M=\sqrt{-\Theta^a \Theta_a}$ satisfies
$0\le \delta\le 2 \pi$ (open universe). In this way we are also able to
establish
whether  $\Theta^a\in V^+$ or $V^-$.
The use of the adjoint representation would
have left us with the ambiguity ($\delta$ ,$\hat n$
)$\leftrightarrow$ ($ 2 \pi -\delta$, $-\hat n$), where $\hat n$ is
the versor  corresponding to the vector defined by (\ref{AAA}). In the
following construction, however, this ambiguity can be avoided by
continuity reasons. On the
other hand for a closed universe the total energy is defined  as the
Euler number of the space 2-manifold.

\section{A positive energy theorem in 2+1 dimension}
We shall assume the existence of an edgeless space-like initial data
two dimensional surface and we shall investigate here the consequences
on the energy of imposing on the matter energy-momentum tensor the
dominant energy condition.
In our analysis we shall use eq.(\ref{E2}) in defining on it the energy.
Given the space-like initial data
two dimensional surface $\Sigma$, we consider on it a family of closed
contours  $x^\mu(s,\lambda)$ with  $0\le \lambda\le 1$  $x^\mu(s,0)=
x^\mu(s,1)$ and such that for increasing $s$  ($s_1<s_2$) the contour
$x^\mu(s_1,\lambda)$ is completely contained in
$x^\mu(s_2,\lambda)$ and which shrinks to a single point for $s=0$.

It is easy to prove (see Appendix A) that the Wilson loop under the
deformation induced  by the parameter $s$ changes according to the
following  equation
\begin{eqnarray}
\label{Evol1}
&&\frac{D W}{d s}(s,1)\equiv \frac{d W }{d s}(s,1)+i [J_a \Gamma^a_{\mu}(s,1)
\frac{d x^\mu}{d s}, W(s,1)]= \nonumber\\
&& i W(s,1)\int^1_0 d\lambda W(s,\lambda)^{-1} R_{\mu\nu}(s,\lambda)
\frac{d x^\mu}{d\lambda}\frac{d x^\nu}{d s} W(s,\lambda)
\end{eqnarray}
where $R_{\mu\nu}$ is the curvature  form in the fundamental
representation. In 2+1 dimensions the curvature 2-form is given
directly by the energy-momentum tensor
\begin{equation}
\label{Moto}
R_{\mu\nu}=8 \pi G ~\eta_{\mu\nu\rho} T^{a\rho} J_a
\end{equation}
where $\eta_{\mu\nu\rho}\equiv \sqrt{- g} \epsilon_{\mu\nu\rho}$ and
$T^{a\rho}$
is the energy momentum tensor contracted with the dreibein, i.e. $T^{a\rho}=
e^a_\mu T^{\mu\rho}$.
 Substituting (\ref{Moto}) into (\ref{Evol1}) we have
\begin{equation}
\label{Evol2}
\frac{D W}{d s}(s,1)= 8 \pi i G~ W(s,1)\int^1_0 d\lambda W(s,\lambda)^{-1}
\eta_{\mu\nu\rho} T^{a\rho}J_a
\frac{d x^\mu}{d\lambda}\frac{d x^\nu}{d s} W(s,\lambda)
\end{equation}
$-\displaystyle{\eta_{\mu\nu\rho}\frac{d
    x^\mu}{d\lambda}\frac{dx^\nu}{d s}}$   is the
area vector $N_\rho$ corresponding to the 2-dimensional surface
$\Sigma$, and thus it is time-like. But then
\begin{equation}
\label{q}
q^a(s,\lambda)\equiv T^{a\rho} N_\rho
\end{equation}
according to the DEC is time-like and future directed. By
substituting eq. (\ref{q}) into eq. (\ref{Evol2}) we have
\begin{eqnarray}
\label{Evol3}
&&\frac{D W}{d s}(s,1)= -8 \pi i G~ W(s,1)\int^1_0 d\lambda W(s,\lambda)^{-1}
J_a q^{a}(s,\lambda)  W(s,\lambda)=\nonumber\\
&&-8 \pi i G ~ W(s,1) J_a Q^a(s)
\end{eqnarray}
where due to the similitude transformation $W^{-1} J_a q^a  W$, also
the integrated $Q^a$ is time-like and future directed.

If we  parametrize $W$ in the fundamental representation as
$W(s,1)=w(s) I - 2 i J_a \theta^a(s)$ the evolution equation gives
\begin{equation}
\label{evolP1}
\frac{d w(s)}{d s}= 4 \pi G  Q_a(s) \theta^a (s)
\end{equation}
and
\begin{equation}
\label{evolP2}
\frac{D \theta^b (s)}{d s}\equiv \frac{ d \theta^b(s)}{d
  s}+\Gamma^b_{c\mu}(s)\frac{dx^\mu}{d s} \theta^c (s)=4\pi G ~(w(s)
Q^b(s)+\epsilon^b_{~lm}\theta^l(s) Q^m(s))
\end{equation}
The  initial condition $W(0,1)=I$ imposes that $w(0)=1$ and
$\theta^b(0)=0$. During the evolution in $s$ the quantity
$w(s)^2-\theta_b(s) \theta^b(s)$ is conserved and the initial
conditions fix its value to $1$.  This simply means that
${\rm det}(W)$ is equal to $1$ for every $s$.

We define $\bar s$ as
the lower bound of the support of $Q^a(s)$, (if the point
($\lambda=0,s=0$) is inside the matter $\bar s$ coincides with
$s=0$, otherwise not). For $s\ge \bar s$,
using the initial condition $w(\bar s)=1$
and $\theta^a(\bar s)=0$,
 from (\ref{evolP1}) and
(\ref{evolP2})  we obtain
\begin{equation}
w(s)=1+4\pi G~\int^s_{\bar s} \theta^a(s^\prime) Q_a(s^\prime)  ds^\prime
\end{equation}
and
\begin{equation}
\label{CCFF}
\theta^a(s)=4\pi G~\int^s_{\bar s} Q^a(s^\prime) ds^\prime+
{\rm higher~~ order~~ in~~} (s-\bar s)^2.
\end{equation}
Being $Q^a(s)$ time-like future-directed also $\theta^a(s)$ is
time-like future-directed in a neighbourhood
of $\bar s$ and then from
eq. (\ref{evolP1}) we have that $w( s)$ is decreasing in the same
positive neighbourhood of $\bar s$.

Now we show that in the range $1\ge w(s)\ge -1$, $w(s)$ is a
decreasing function of $s$ and
\begin{equation}
\label{dismassa}
w(s)\le \cos \left (4 \pi G\int_0^s  \sqrt{-Q^a(s^\prime)Q_a(s^\prime)}
d s^\prime\right )
\end{equation}
i.e. $M(s)\ge \int_0^s \sqrt{-Q^a(s^\prime) Q_a(s^\prime)}~ds^\prime$,
where $M(s)=\frac{1}{8\pi G}
\sqrt{-\Theta^a(s)\Theta_a(s)}$ is the total mass of the
subsystem enclosed by the loop $x^\mu(s,\lambda)$, $0\le \lambda\le
1$.

In fact from eq. (\ref{evolP1}) we have for $-1\le w(s)\le 1$,~~
$-\theta^a(s)\theta_a(s)\ge 0$ i.e. $\theta^a(s)$ is time-like and
being initially future-directed remains future directed and thus
\begin{equation}
-\theta^a(s) Q_a(s)=Q^0(s)\theta^0(s) -\bftheta (s)\cdot{\bf Q}(s)>
m(s) \theta(s)
\end{equation}
with $m(s)=\sqrt{-Q^a(s)Q_a(s)}$ and
$\theta(s)=\sqrt{-\theta^a(s)\theta_a(s)}=\sqrt{1-w(s)^2}$. Using
this inequalities eq. (\ref{evolP1}) becomes
\begin{equation}
\label{dis1}
\frac{d w(s)}{d s}\le- 4 \pi G m(s)\sqrt{1-w(s)^2}\le 0
\end{equation}
from which we see that for $-1\le w(s)\le 1$, $w(s)$ is a decreasing
function of s.
%Thus if there exists a point $\bar s$ in which
%$w(\bar s)=1$ and $w(s)$ decreasing in the positive neighbourhood of
%$\bar s$, the function $w(s)$ is monotonic until reaches the value
%$-1$. Now we prove that such a point exists.

Physically the monotonic decrease of $w(s)$ from $1$ to $-1$ means
that by adding matter satisfying the DEC the mass of the system
increases up to its final value which for an open universe is less
than  $\displaystyle{\frac{1}{4 G}}$, i.e. $\delta\le 2\pi$ and thus
the final value of $w$ is larger than $-1$. From this we have also for an
open universe that $\theta^a(s)$ and thus $\Theta^a (s)$ is a future
directed  time-like vector. In conclusion for an open universe with
matter sources satisfying the DEC the total energy-momentum vector
defined by (\ref{E2}) is time-like.

To prove eq. (\ref{dismassa}) we write using $-1\le w(s)\le 1$, $w(s)=
\cos(f(s))$ with $0\le f(s)\le \pi$. Substituting in the
inequality (\ref{dis1}) we have
\begin{equation}
-\sin(f(s)) f^\prime(s)\le -4\pi G m(s)~ \sin (f(s))
\end{equation}
But then
\begin{equation}
f^\prime (s)\ge 4 \pi G~ m(s)\longrightarrow f(s)\ge 4\pi G~\int^s_0
m(s^\prime) d s^\prime
\end{equation}
from which (\ref{dismassa}) follows.
Physically such a relation means that for an open universe with matter
satisfying the DEC the total mass of the universe is higher or equal
to $\int^\infty_0 m(s^\prime) d s^\prime$, a result which is common to
Minkowski space, despite the non linear composition law of momenta in
2+1 gravity. Such inequality is optimal in absence of the other
hypothesis, as  it is saturated by the static universe \cite{DJH}.
We want now to inquire what happens if $w(s)$ instead of stopping in
its evolution before reaching $-1$, reaches this value at $s=s_0$.
In $w(s_0)=-1$,  $\theta^a(s_0)\theta_a(s_0)=0$ which means that
$\theta^a (s_0)$ is either the zero vector or a light-like vector.
First we consider  the zero vector alternative. From (\ref{evolP2}) we
see that for $s=s_0+\epsilon$,  $\theta^a(s)$ is a past-directed
time-like vector, thus giving $w(s_0+\epsilon)>-1$. With identical
reasoning as before one can prove that $w(s)$ is monotonically
non-decreasing function until it reaches the value 1 which correspond
to the closure of the universe because in this case
the total mass of the universe
is $1/2 G$ (i.e. $\delta=4\pi$). Even if $\theta_a(s)$ for $ s>s_0$ is
time-like past-directed, the energy-momentum vector of our subsystem
(contained  in the contour $s$)
\begin{equation}
\Theta^a(s)=\frac{M(s)}{\sin M(s)/2} \theta^a(s)
\end{equation}
is future-directed because now $\displaystyle{\sin\frac{M(s)}{2}<0}$.
We come now to the case in which $\theta_a(s_0)\not =0$ and
light-like, necessarily future directed because the limit of a
time-like  future directed vector. From eq. (\ref{evolP1}) we have
that $w(s_0+\epsilon)<-1$ and thus  $\theta_a(s_0+\epsilon)$ is
space-like. The simplest example of such a situation is the Gott-pair
in which two  particles each  having time-like  momentum give rise to
a universe with total space-like momentum \cite{DJH2}. We recall that for such
system there exists space-like surfaces \cite{DJH2,Cut}
which satisfy the condition of our theorem.

Adding matter satisfying the DEC we cannot return to an open universe
(with $\delta<2 \pi$). In fact in such a circumstance   there is
a value of $s$, $s_1>s_0$, where $w(s_1)=-1$ and it is increasing.
At $s=s_1$ we  also have either $\theta^a(s_1)\in L^+$ or
$\theta^a(s_1)=0$. For $\theta^a(s_1)\in L^+$,  $Q_a \theta^a(s_1)<0$
giving $\displaystyle{\frac{dw}{ds}<0}$ which contradicts the fact that
$w(s)$ increases at $s=s_1$.   For $\theta_a(s_1)=0$  we have
$w(s)=-1-8\pi^2 G^2 Q_a(s_1) Q^a(s_1) (s-s_1)^2+{\rm o}((s-s_1)^2)$ which
contradicts again the assumption that  $w(s)$ increases at $s=s_1$.
On the other hand by properly adding matter one can land to universe
with $M(s)>1/4G$. This is achieved when $\theta_a(s_1)\in L^-$. In fact
now  from (\ref{evolP1}) ~~$\displaystyle{\frac{dw}{ds}>0}$, thus $w(s)$
increases above $-1$ and $\theta^a(s)\in V^-$ which  means that the mass
has increased above $2\pi$.
This theorem provides a generalization of the theorem \cite{CFGO} stating that
for an open universe  composed by a collection of spinless point
particles, with total time-like momentum, no subsystem can possess
space-like momentum. Here the proof is  given for any  distribution of
matter  satisfying the DEC.
As final check of our definition of energy we show how in this formalism
one can recover the usual ansatz for the static problem.
We notice that in the  static case  the dreibeins can be chosen
independent of time and of the form
$e^a_{~0}= N \delta^a_0$, $e^0_{~i}=0$ and
we have \cite{DJH} for the energy-momentum
tensor $T^{00}= N {\cal T}^{00}$ where ${\cal T}^{00}$
is the $00$ component of the energy-momentum tensor in the base given by the
coordinates.
Substituting into eq. (\ref{Evol1}) we have
\begin{equation}
\frac{D W}{d s}(s,1)= 8 \pi i G~ W(s,1)\int^1_0 d\lambda W(s,\lambda)^{-1}
\sqrt{-g} N { \cal T}^{00}J_0
\left(\frac{d x^1}{d\lambda}\frac{d x^2}{d s}-
\frac{d x^2}{d\lambda}\frac{d x^1}{d s}\right )
 W(s,\lambda)
\end{equation}
which is solved by
\begin{equation}
W=V^{-1} (s)\exp(-8\pi iG J_0\int {\cal T}^0_0 \sqrt{-\gamma} dx^1 dx^2) V(s)
\end{equation}
where $\gamma$ is the determinant of the space metric and $V(s)$ is the
holonomy of $\Gamma^a_s(s,1)$ computed along $x^\mu(s,1)$. Then we have
\begin{equation}
\delta=8\pi G M= 8\pi  G \int{\cal T}^0_0 \sqrt{-\gamma} dx^1 dx^2=
4\pi  G \int R^{(2)} \sqrt{-\gamma} dx^1 dx^2
\end{equation}
in agreement with the result of \cite{DJH}.

\typeout{inizio}

\section{Poincar\'e group and a definition of angular momentum}

This section is devoted to an invariant definition of the angular momentum
of the system by means of the Wilson loop. To this purpose  we introduce
the complete holonomy associated to the $ISO(2,1)$  group i.e.
\begin{equation}
\label{*}
W={\rm Pexp}\left ( -i \oint J_a\Gamma^a_\mu dx^\mu+ P_a e^a_\mu dx^\mu\right )
\end{equation}
computed on a loop that encloses all matter, e.g. a loop at space
infinity.
%In $2+1$ dimensions this object is
%meaningful because Einstein theory can be viewed as the Chern-Simon theory
%of $ISO(2,1)$  (Witten).
%Suppose now that we are able to calculate the r.h.s.
%of eq. (\ref{*}) and  rewrite it as
Such a loop (\ref{*}) can always be written  in the form
\begin{equation}
W=\pm {\rm exp}\left (-i J_a \Theta^a-iP_a \Xi^a\right )
\end{equation}
As shown in Appendix B, $\Theta^a$ is the total energy-momentum defined
in the previous
sections. Now we shall see that it is natural to identify $\Xi^a$ with
the total angular
momentum of the system; however this identification requires some caution.
Contrary to what happens for $\Theta^a$, which, up to a Lorentz transformation,
is independent of the loop  chosen to calculate it, $\Xi^a$ and its modulus
depend  on the point $O$ where the loop closes.
%and on the loop itself.
(See Appendix B for a complete treatment of the transformation law
of $\Xi^a$). This fact is neither  contradictory nor unexpected if we recall
what happens to the total angular momentum of a mechanical system in
special relativity \cite{L-L}. Also in that
case it depends both on the origin of the reference frame and on the frame
itself. The only quantity, which is invariant and meaningful,
is the component of the angular momentum  along the total momentum. In our
formalism this quantity is
\begin{equation}
\label{JJ}
{\cal J}=-\frac{\Theta_a \Xi^a}{8\pi G\sqrt{-\Theta_a\Theta^a}}
\end{equation}
and, as expected, it does  not depend on the loop used
to compute it.
%In fact it corresponds to a Casimir of $ISO(2,1)$.
The
proof of the invariance of $\cal J$ defined by two equivalent loops can be
obtained using the results of Appendix B.
In the following when we speak about angular momentum we refer to the
definition (\ref{JJ}). The first check for our definition  is to compute it
for the geometry of a localized source (cosmic string ).
Using for
convenience as closed contour a circle of radius $R$ on the surface
$t={\rm const}$
and keeping in mind that the metric is independent of the angular  variable
$\theta$ we obtain
\begin{equation}
W={\rm exp}\left (-i 2\pi J_a \Gamma^a_\theta
-2 \pi i P_a e^a_\theta \right )
 \end{equation}
Using  the explicit expression of $e^a_\theta$ and $\Gamma^a_\theta$
for a
cosmic string we obtain
\begin{equation}
\label{RRR}
{\cal J}=-\frac{\Gamma^a_\theta e_{a \theta}}{8\pi G \sqrt{\Gamma^a_\theta
\Gamma_{a\theta}}}=J
\end{equation}
where $J$ is the usual angular momentum defined by the time-shift
which appears in a syncronous coordinate system when one encircles the source
\footnote{We notice that eq. (\ref{RRR}) can be considered a simple formula
 which gives the angular momentum  for all axially symmetric geometries, once
expressed in a reference in which the metric is independent of $\theta$.}.
Coming back to the general problem, also in this case we are able to write an
evolution equation for the Wilson loop under a deformation of the contour.
Generalizing the argument of  the previous section we obtain
\begin{eqnarray}
&&{D W\over ds}={d W\over ds}+i[\Gamma_s,W]+i[e_s,W]=
\nonumber\\
&&=iW(s,1)\int_0^1 d\lambda {W(s,\lambda)}^{-1} \{ J_a R^a_{\lambda
s}(s,\lambda)+P_a S^a_{\lambda s}(s,\lambda)\}W(s,\lambda)
\end{eqnarray}
where we have projected the group curvature on the usual geometric curvature.
Taking into account that our theory is torsionless and that
\begin{equation}
R_{\mu\nu}=8 \pi G ~\eta_{\mu\nu\rho} T^{a\rho} J_a
\end{equation}
we obtain
\begin{equation}
\label{OOO}
\frac{D W}{d s}(s,1)= -8 \pi i G~ W(s,1)\int^1_0 d\lambda W(s,\lambda)^{-1}
J_a q^{a}(s,\lambda)  W(s,\lambda)
\end{equation}
where $q^a(s,\lambda)$ is the same quantity defined in sec. 3. The substantial
difference between this equation and eq. (\ref{Evol3}) is that now $W$ are
Poincar\'e transformations. This prevents us from rewriting the  integral in
eq. (\ref{OOO}) simply as a future directed time-like vector $Q^a(s)$.
In fact the translation part of $W(s,\lambda)$ generates a new term which
has no particular  properties  even when we impose that our energy-momentum
tensor satisfies the DEC. In detail, recalling the structure of the adjoint
representation of Poincar\'e group,  we get
\begin{equation}
\label{MMMAAA}
\frac{D W}{ds}=-8\pi i G~W \int^1_0 d\lambda (J_a A^a_l(s,\lambda)
q^l(s,\lambda)+
P_a l^a(s,\lambda))=-8\pi i G W(J_a Q^a+ P_a L^a)
\end{equation}
where $Q^a(s)$ is the  same vector as the one defined in the previous
section and $l^a$ is given by
\begin{equation}
\label{DFG}
l^a(s)= \epsilon^a_{~bc} A^b_l(s,\lambda) q^l(s,\lambda) \Xi^c(s,\lambda)
\end{equation}
where $A^b_l(s,\lambda)$ is thge matrix action for
the operator $\exp(iJ_a \Theta^a(s,\lambda))$ in the adjoint representation.
To write
explicitly eq. (\ref{MMMAAA}) we introduce the following $4\times4$
representation
\begin{equation}
\label{POIN}
J_a= \left (
\begin{array}{cc}
 I_a  &   0\\ 0 &  I_a
\end{array} \right )
{}~~~~~P_a=\left(
\begin{array}{cc}
0 & I_a\\
0 &   0
\end{array}
\right)
\end{equation}
where $I_a$ are the generators of $SO(2,1)$ in the fundamental representation.
It is easy to verify that the matrix defined in eq. (\ref{POIN}) satisfy the
algebra
of $ISO(2,1)$ and in addition $P^a$ is nihilpotent. Now we parametrize the
generic transformation in this representation as follows
\begin{equation}
W=w(s)- 2 i J_a \theta^a (s)+ u(s) K-2 i P_a v^a(s)
\end{equation}
where
\begin{equation}
K=\left (\begin{array}{cc}
0 & 1\\
0 & 0
\end{array}\right ).
\end{equation}
Substituting the previous parametrization and the explicit form of generators
in eq. (\ref{MMMAAA}) we recover the same equations of sec. 3 for the sector
regarding the Lorentz transformations, and two new equations for the sector
regarding the translations
\begin{eqnarray}
\label{DDD}
&&\frac{d u(s)}{ds}=4\pi G( v_a(s) Q^a(s)+\theta^a(s) L_a(s))\nonumber\\
&&\frac{D v^a(s)}{ds}=4\pi G (u(s) Q^a(s) +\epsilon^a_{~bc} v^b(s) Q^c(s)+
w(s) L^a(s)+ \epsilon^a_{~bc} \theta^b(s) L^c(s)).
\end{eqnarray}
Eq. (\ref{DDD}) relate the angular momentum to the spacial distribution of
energy and momentum. Here we shall use eq.(\ref{DDD}) to prove that for weak
sources the given definition of angular momentum identifies with usual
definition of special relativity\cite{L-L}.
In fact to first order writing explicitly the covariant derivative we
have
\begin{equation}
\label{NBN}
{d v^a(s)\over ds}=4\pi G  L^a(s) + \epsilon^a_{~\mu b}{dx^\mu(s,1)
\over ds}\theta^b(s)
\end{equation}
and form eq. (\ref{evolP2})  with the same degree of approximation
\begin{equation}
\label{q1}
\theta^b(s)= 4\pi G \int^s_0 ds' \int^1_0d\lambda~ q^b(s',\lambda).
\end{equation}
Moreover we notice that in eq.(\ref{MMMAAA})
\begin{equation}
\label{qq}
L^a(s)=\int \epsilon^a_{~\mu b}(x(s,\lambda) -x(s,0))^\mu
q^b(s,\lambda)) d\lambda
\end{equation}
because to first order in eq. (\ref{DFG}) $\Xi^a(s,\lambda)=x^a(s,\lambda)-
x^a(s,0)$ and the rotation matrix can be taken equal to the identity.
We now substitute eqs. (\ref{q1}) and (\ref{qq}) into eq. (\ref{NBN})
and integrate in $s$. Integrating by
parts the $\theta$ term we obtain
\begin{equation}
\Xi^a(s)\approx 2 v^a(s)= -
8\pi G \int_0^s ds' \int_0^1 d\lambda\epsilon^a_{~\mu b}
{}~x^\mu(s^\prime,\lambda) q^b(s',\lambda)+
2\epsilon^a_{~\mu b} x^\mu(s,0) \theta^b(s)
\end{equation}
If now we take into account
 the relation between $q^a(\lambda,s)$ and the energy-momentum
tensor and the eq. (\ref{CCFF}) to the lowest order in $G$
the previous formula can be rewritten as
\begin{equation}
\Xi^a(s)\approx 2 v^a(s)= -
8\pi G  \int dx^1 dx^2  \epsilon^a_{~\mu \nu}
 x^\mu {\cal T}^{\nu 0}+
8\pi G\epsilon^a_{~\mu \nu} x^\mu(s,0) \int dx^1 dx^2 {\cal T}^{\nu 0}
\end{equation}
which is $-8\pi G$ times the expression of
 angular momentum in special relativity
\cite{L-L}. The last term means that we are not computing the angular
momentum with respect to the origin but with respect to the point  where
the loop  closes. This can be easily understood if we keep in mind that
only this  point is strictly related to the holonomy.

%We are not able to discuss this equation. But they are also interesting
%because they express a relation between the mass distribution and angular
%momentum.
%Qui tu eventualmente inserire altri commenti se vuoi ......
\typeout{fine}

\section{Anti de Sitter 2+1 dimensional gravity}

The problem of defining the energy in de Sitter and anti de Sitter
3+1 dimensional gravity has been treated by Abbot and Deser \cite{DesAb}, with
the result that it is possible to define an energy associated to the
generator $J_{04}$, which is de Sitter rather that Poincar\'e covariant.
The results of sect.3 can be easily extended to anti de Sitter 2+1
dimensional gravity due to the fact that the anti de Sitter algebra is
the algebra of $SU(1,1)\times SU(1,1)$.
In this case it is  meaningful  to consider
only the Wilson loop the full connection $J_a\Gamma^a+{1\over
l}P_a e^a$, because the loop of the $J_a\Gamma^a$ does not
generate invariants under the full anti de Sitter group.
We adopt the following Lie algebra
\begin{equation}
[J_a,J_b]=i\varepsilon_{abc}J^c
\end{equation}
\begin{equation}
[J_a,P_b]=i\varepsilon_{abc}P^c
\end{equation}
\begin{equation}
[P_a,P_b]= {i l^2}\varepsilon_{abc}J^c.
\end{equation}
Following the procedure of the previous
section we derive the evolution equation for the Wilson loop
$$
{D W\over ds}\equiv{d W\over ds}+i[\Gamma_s,W]+i[e_s,W]=
$$
\begin{equation}
\label{EEE}
iW(s,1)\int_0^1 d\lambda {W(s,\lambda)}^{-1} \{J_a R^a_{\lambda
s}(s,\lambda)+P_a S^a_{\lambda s}(s,\lambda)+
i[e_\lambda(s,\lambda)),e_s(s,\lambda)]\}W(s,\lambda)
\end{equation}
where $\Gamma=J_a\Gamma^a$ and $e=P_a e^a$ and $R^a_{\lambda s}$ is
the usual curvature and $S^a_{\lambda s}$ is the
torsion which in the following will be set to $0$.

Using Einstein's equations
\begin{equation}
R_{\mu\nu}+i[e_\mu,e_\nu]=8\pi G\eta_{\mu\nu\rho}T^{a\rho}J_a
\end{equation}
we can rewrite eq.(\ref{EEE}) as
\begin{equation}
{DW\over ds}=-8\pi i G W(s,1)\int_0^1d\lambda {W(s,\lambda)}^{-1}
q^a(s,\lambda) J_a W(s,\lambda)
\end{equation}
Defining in the anti de Sitter case
\begin{equation}
J^{\pm}_a=\frac{1}{2}\left (J_a\pm{1\over l}P_a\right )
\end{equation}
we have that
\begin{equation}
[J^{\pm}_a,J^{\pm}_b]=i\epsilon_{abc}J^{\pm c}
\end{equation}
while
\begin{equation}
[J^{+}_a,J^{-}_b]=0.
\end{equation}
Using these generators the previous equation separates into two
independent equations
\begin{eqnarray}
{DW^{\pm}\over ds}&&=-8\pi i G W^\pm(s,1)\int_0^1d\lambda~
{W^\pm}(s,\lambda)^{-1}
q^a(s,\lambda) J^\pm_a W^\pm(s,\lambda)=\nonumber\\
&&-8\pi i W^\pm(s,1)J^\pm_a Q^{\pm a}(s)
\end{eqnarray}
where
\begin{equation}
W^\pm={\rm Pexp}\left ( -i\int J^\pm_a(\Gamma^a_\mu\pm l e^a_\mu)dx^\mu
\right ).
\end{equation}
We remark that $Q^{\pm a}(s)$ are future directed time like vectors
because $W^{\pm}$ induce a similitude transformation which is a real
Lorentz transformation on $Q^a$. Parametrizing $W^\pm$ in the
fundamental representation as follows
\begin{equation}
W^\pm=e^{-i J^\pm_a \Theta^{\pm a}}=w^\pm-2iJ^\pm_a\theta^{\pm a}
\end{equation}
we obtain the equations
\begin{equation}
{dw^\pm(s)\over ds}=4\pi GQ^\pm_a(s)\theta^{\pm a}(s)
\end{equation}
and
\begin{eqnarray}
{D\theta^{\pm a}(s)\over ds}=&&{d\theta^{\pm a}(s)\over
ds}+(\Gamma^a_{b\mu}\pm l\epsilon^a_{~bc}e^c_\mu)\theta^{\pm b}(s)
{dx^\mu(s,1)\over ds}=\nonumber\\
&&4\pi G (w^\pm(s) Q^{\pm a}(s)+\epsilon^a_{~lm}\theta^{\pm l}(s)Q^{\pm m}(s)).
\end{eqnarray}
We notice that $w^\pm$ and $\theta^\pm$ obey the same initial
conditions as in the Poincar\'e case and except for the term $\pm l
\epsilon^a_{~bc}e^c_\mu\theta^{\pm b}(s)\displaystyle{\frac{dx^\mu(s,1)}{ds}}$
the same equations. As the
additional term contains the $\epsilon^a_{~bc}$ we still have as in the
Poincar\'e case $w^{\pm~2}-\theta^{\pm~2}=1$.

We can now carry over exactly the same discussion given in Sec.III to
the vectors $\theta^\pm$.  Again the conclusion is that $w^\pm$ which
are initially equal to 1, as matter is included in the loop, decrease
below 1 while $\theta^\pm$ become time-like future directed vectors.
The monotonic decrease of $w^\pm$ goes on until they reach the value
-1. Thus from eq.(62) we see that for $s<s_0^+,~ s<s_0^-$ i.e. before
any of the two $w^\pm$ reaches -1, the  mass of the system is
monotonic non decreasing. If above $s_0^\pm$ both $\Theta^{+a}$ and
$\Theta^{+a}$ remain time-like we have still a monotonic increase of
mass. Instead if one of the two vectors at the critical point goes over
to a space-like vector, the situation is no longer clear cut because we
cannot decide a priori whether the square of the invariant mass
remains positive.

The usual mass $M$ and angular momentum $J$ can be recovered by the
following two combinations of the invariants associated to $\Theta^\pm_a$
\begin{equation}
M=\frac{1}{8\pi G}\sqrt{-\frac{\Theta^{+a}\Theta^+_a+\Theta^{-a}\Theta^-_a}{2}}
\end{equation}
and
\begin{equation}
{\cal J}=-\frac{1}{256 \pi^2 G^2 l M}
(\Theta^{+a}\Theta^+_a-\Theta^{-a}\Theta^-_a)
\end{equation}
If the cosmological constant goes to zero, these equations becomes the
definitions given in the previous section for the Poincar\`e group.
In the region where the  deficit angles,
 associated to $\Theta^+_a$ and $\Theta^-_a$,
are less then $2\pi$  the following inequality between mass
and angular momentum holds
\begin{equation}
\mid {\cal J}\mid \le M/2 l
\end{equation}
where $l^2$ is the cosmological constant. This fact allows the existence of
black-holes in the anti de Sitter gravity \cite{Teitelboim}.

\section{Conclusions}
Due to the angular deficit at infinity which appears in 2+1
dimensional gravity, it is not possible to carry over from gravity in
$3+1$ dimensions the usual
procedure for defining energy and momentum. The holonomies provide an
invariant, consistent procedure
for defining the total energy and angular momentum for a system in
$2+1$ dimensions. They give rise to conserved quantities which for
small masses go over to the minkowski results.
Given an initial value space-like surface, we considered a family of
widening loops and proved that for any matter energy-momentum tensor
that satisfies the dominant energy condition, the energy-momentum of
the subsystem enclosed by the loop is a time-like future directed
vector, whose norm increases as the loop widens.

By enclosing more and more matter satisfying the DEC, one reaches the
angular deficit of $2\pi$ and then one can either go over to a
time-like universe which, having an angular deficit greater that
$2\pi$ closes kinematically, or to a space-like universe of the Gott
type.

By adding more matter one can never come back to a space-like open
universe; this provides an extension of the result by Carroll, Fahri,
Guth and Olum to the most general matter energy-momentum tensor
satisfying the DEC, i.e. that in 2+1 dimensions if a
subsystem has space-like momentum the universe is either closed or
space-like.

Without much changes the above mentioned results are carried over to
the anti de Sitter case.

\section{Acknowledgements}
We are grateful to D. Bak, D. Cangemi, S. Deser and  R. Jackiw,
  for useful discussions  and
one of us D.S. thanks Della Riccia Foundation for financial support.
\appendix
\section{}
 In this appendix we derive the equation for the change of a
Wilson loop under a continuous deformation.
Consider a family of curves parametrized by a parameter $\lambda$
which describes the curve and a parameter $s$ which labels the member
of the family of curves. Given a connection $A(x)$ on the surface
described by the two parameters $\lambda$ and $s$ we can consider the
two components $A_\lambda$ and $A_s$ of the connection according to
the formulae
\begin{equation}
A_\lambda(s,\lambda)=A_\mu(x){dx^\mu\over d\lambda}~~~~{\rm and}~~~~
A_s(s,\lambda)=A_\mu(x){dx^\mu\over d s}
\end{equation}
The path ordered integral along the generic arc is given by
\begin{equation}
W(s,\lambda)={\rm Pexp}(-i\int_0^\lambda A_\lambda(s,\lambda')d\lambda')
\end{equation}
One now considers the derivative of $W(s,\lambda)$ with respect to
$s$; actually it is more significant to consider the so called covariant
derivative of $W(s,\lambda)$ given by
$$
{D W(s,\lambda)\over \partial s}={\partial W(s,\lambda)\over \partial
s} + i(A_s(s,\lambda) W(s,\lambda) - W(s,\lambda)A_s(s,0))=
$$
\begin{equation}
\label{IIIII}
={\partial W(s,\lambda)\over \partial
s} + iW(s,\lambda)(W(s,\lambda)^{-1} A_s(s,\lambda) W(s,\lambda) -A_s(s,0))
\end{equation}
The first term on the r.h.s. of (\ref{IIIII})is given  by \cite{VolDol}
\begin{equation}
\label{VVV}
-iW(s,\lambda)\int_0^\lambda d\lambda' W(s,\lambda)^{-1}{\partial
A_\lambda(s,\lambda')\over \partial s} W(s,\lambda')
\end{equation}
On the other hand the second term in (\ref{IIIII}) can be written as
\begin{equation}
iW(s,\lambda)\int_0^\lambda{\partial \over \partial \lambda'}
(W(s,\lambda')^{-1}A_s(s,\lambda')W(s,\lambda'))d\lambda'
\end{equation}
because $W(s,0)=I$. Performing the derivative in the previous equation
and adding to eq. (\ref{VVV}) we have
\begin{eqnarray}
\label{A7}
{D W(s,\lambda)\over \partial s}&&=-iW(s,\lambda)\int_0^\lambda d\lambda'
W(s,\lambda')^{-1}\biggl [{\partial A_\lambda(s,\lambda') \over \partial s}-
{\partial A_s(s,\lambda') \over \partial \lambda}+\nonumber\\
&&i A_\lambda(s,\lambda') A_s(s,\lambda')-i
A_s(s,\lambda')A_\lambda(s,\lambda')
\biggr ]
W(s,\lambda')=\nonumber\\
&&-iW(s,\lambda)  \int_0^\lambda   d\lambda'
W(s,\lambda')^{-1} F_{s\lambda}(s,\lambda')   W(s,\lambda')
\end{eqnarray}
Obviously these equations are valid in $n$ dimensions. The simplifying
feature in $2+1$ dimensions, that is exploited in the text, is that one can
express the curvature directly in terms of the matter energy-momentum tensor.

\section{}
In this appendix we discuss the transformation properties of the
energy momentum vector and the angular momentum, defined in sec. 2, under
the change of origin of the loop.
 In the following we give the complete treatment
 for the Poincar\'e case, however all the results can be easily
extended to  anti de Sitter case $SO(2,2)$.

\noindent
Given a closed contour
$\Lambda$ we consider two points  $O$ and $O^\prime$ on it. The Wilson loop
$W_O$ with origin $O$ is related to that with origin $O^\prime$
by the following rule
\begin{equation}
\label{eqw}
W_O=V^{-1}_{O^\prime O}~ W_{O^\prime}~ V_{O^\prime O}
\end{equation}
where $V_{O^\prime O}$ is the gauge transformation associated to the path
from $O$ to ${O^\prime }$ along the loop $\Lambda$.
Due to eq. (\ref{eqw}) $W_O$ and $W_{O^\prime}$ have the same secular
polynomial and so possess the same invariants. In other words the two
loops
define the same total mass and component of angular momentum along the
total momentum.
The explicit form of
$V_{O^\prime O}$ is given by the following path ordered integral
\begin{equation}
V_{O^\prime O}={\rm Pexp}\left (- i~\int^{O^\prime}_O J_a \Gamma^a_\mu  dx^\mu+
P_a e^a_\mu  d x^\mu \right).
\end{equation}
We can explicitly calculate the sector regarding the translation
generators because they commute among themselves and we obtain the
 formula
\begin{equation}
V_{O^\prime O}={\rm Pexp}\left (- i~\int^{O^\prime}_O J_a \Gamma^a_\mu  dx^\mu
\right ) \times {\rm exp}\left (-i \int^{O^\prime}_O P_a T^a_l
(\lambda) e^l_\mu dx^\mu \right )
\end{equation}
where $T^a_l=[{\rm Pexp}\left (- i~\int^{P}_O J_a \Gamma^a_\mu dx^\mu\right
)]^{-1}$ in the adjoint representation  and $P$ is a generic point between
$O$ and $O^\prime$.
This result can be easily found using the formula  of \cite{VolDol} which
relates the Pexp-integral of the sum $A+B$ of two quantities $A$ and
$B$ to the
Pexp-integral of $A$ and that of $B$ and recalling that $[ P_a, P_b]=0$. For
brevity we define
\begin{equation}
\label{B4}
S^a=\int^{O^\prime}_O T^a_l (\lambda) e^l_\mu  dx^\mu ~~~~~~~~{\rm and}~~~~~~~~
A={\rm Pexp}\left (- i~\int^{O^\prime}_O J_a \Gamma^a_\mu dx^\mu\right )
\end{equation}
and we rewrite
\begin{equation}
\label{HHH}
V_{O^\prime O}=A\times {\rm exp}\left ( -i P_a S^a \right)
\end{equation}
This formula has a simple geometrical interpretation: $S^a$ is the
shift between
the two origins and $A$ is the Lorentz transformation which relates
the dreibein of the observer in $O$ with that of the observer in $O^\prime$.

\noindent In the following we consider the case in which our Wilson
loop exponentiates, i.e. it can be parametrized as
\begin{equation}
\label{PPP}
W_O={\rm exp}(-i J_a \Theta^a_O -i P_a \Xi^a_O ),
\end{equation}
and the same parametrization  holds for $W_{O^\prime}$. Substituting
eq. (\ref{PPP}),  its analogous for $W_{O^\prime}$ and eq. (\ref{HHH})
into eq. (\ref{eqw})
we get the following transformation rules for $\Theta^a$ and $\Xi^a$
\begin{equation}
\label{TT1}
\Theta^a_{O^\prime}= A^a_l \Theta^l_O
\end{equation}
\begin{equation}
\label{TT2}
\Xi^a_{O^\prime}=A^a_l (\Xi^l_O+\epsilon^l_{~~bc} S^b\Theta^c_O ).
\end{equation}
where $A^a_l$ is the second operator in eq. (\ref{B4}) in the adjoint
representation.
If we recall the classical transformation law for the energy-momentum vector
and angular momentum under a change of the origin and a rotation of  reference
 frame, we see that they have the same form of eq.(\ref{TT1}) and
eq.(\ref{TT2}). This fact is a further hint for
identifying $\Theta^a$ and $\Xi^a$ with energy-momentum vector
and angular momentum  of the system.

We consider also the relation between the $\Theta^a$ and $\Xi^a$ for two
equivalent loops, i.e, two loop which can be
deformed one into the other without crossing the matter. The loop are
similar in this case too, i.e.
\begin{equation}
\label{QQQ}
W=V^{-1} W^\prime V
\end{equation}
This is a well known  fact, but one can show it also using eq. (\ref{A7}) of
appendix A. It gives
\begin{equation}
\label{FFF}
\frac{D\hat W}{ds}\equiv \frac{\partial \hat W}{\partial s}-i [\hat W, e_s]-
i [\hat W, \Gamma_s]=0
\end{equation}
where $\hat W(0)=W$ and $\hat W(1)=W^\prime$. This equation gives exactly
the relation (\ref{QQQ}) between the two loops with
\begin{equation}
\label{B11}
V={\rm Pexp}\left (-i \int^1_0 ds (J_a \Gamma^a_s +  P_a e^a_s )\right ).
\end{equation}
where in (\ref{B11}) the integral runs along the trajectory of the origin of
the loop as $s$ varies.
The structure of $V$ is completely analogous to that of $V_{O^\prime O}$,
which implies that the transformation laws between the total momentum
and angular momentum, defined by the two loops, are identical to those
found in the previous case.


\begin{references}
\bibitem{ADMLW}  Einstein A. {\it S.B. preuss.Akad.Wiss.}  (1918) 448;
Arnowitt R., Deser S. and  Misner C.{\it Nuovo Cimento}
{\bf 19} (1961) 668; Landau L. and Lifshitz E. {\it The
Classical Theory of Fields}, Addison-Wesley Press, 1951, p. 316; Weinberg
S. {\it Gravitation and Cosmology}, John Wiley \& Sons, 1972, New York.
\bibitem{DJH} Deser S., Jackiw R. and `t Hooft G.,{\it Ann. Phys.
(N.Y.)}  {\bf 152} (1984) 220.
\bibitem{Witten} Achucarro, A. Townsend, P.K. {\it Phys. Lett.} {\bf B180}
(1986) 89;Achucarro, A. Townsend, P.K. {\it Phys. Lett.} {\bf B229}
(1989) 383; Witten E. {\it Nucl. Phys} {\bf B311} (1988) 46.
\bibitem{CFGO} Carroll S. M., Farhi E., Guth A. H. and Olum K. O.
{\it`` Energy-momentum tensor restrictions on the creation of Gott-time
machine''}, CTP \# 2252, gr-qc/9404065.
\bibitem{HawEll} Hawking S. and Ellis G. {\it The large structure of
space-time}, Cambridge Univ. Press., 1973, pg. 49.
\bibitem{BCJ} Bak D., Cangemi D. and Jackiw R.  {\bf 49} (1994) 5173.
{\it Phys. Rev. D}.
\bibitem{Gott} Gott J.R., {\it Phys. Rev. D}  {\bf 46} (1991) 1126.
\bibitem{DJH2} Deser S., Jackiw R. and `t Hooft G. {\it  Phys. Rev. Lett.}
{\bf 68} (1992) 2647.
\bibitem{Cut}   Cutler C. {\it Phys. Rev. D} {\bf 45} (1992) 487.
\bibitem{L-L} Landau L. and Lifshitz E. {\it The Classical Theory of Fields},
Addison-Wesley Press, 1951, p. 37.
\bibitem{DesAb} Abbot L.F.  and Deser S. {\it Nucl. Phys.} {\bf B195} (1982)
76.
\bibitem{Teitelboim} Banados M., Teitelboim C. and Zanelli J. {\it Phys. Rev.
Lett.} {\bf 69} (1992) 1849; Banados M., Hennaux M, Teitelboim C. and
Zanelli J. {\it Geometry of the (2+1) black hole}, IASSNS-HEP-92-81,
gr-qc/9302012; Cangemi D., Leblanc M. and Mann R.B. {\it Phys. Rev. }
{\bf D48} (1993) 3606.
\bibitem{VolDol} Volterra V. and Hostinsky B. {\it Operations infinitesimale
lineaires},  Gauthiers Villars, 1939, Paris;  Dollard J.D.   and Friedman C.N.
{\it Product integration with applications to differential equations},
Addison Wesley Pub. Co., 1979, Reading.
\end{references}
 \end{document}